\documentclass[%
 reprint,
 amsmath,
 amssymb,
 aps,
]{revtex4-1}
	\usepackage{array}
	\usepackage{enumitem}
	\usepackage{amsthm}
	\usepackage{xcolor}
	\usepackage{amssymb}
	\usepackage{graphicx}
	\usepackage{hyperref}
	\usepackage{mathtools}
	\usepackage{booktabs}
	\usepackage{lineno}
\setlist[description]{leftmargin=\parindent, labelindent=\parindent}

\begin{document}
\title{Buffering variability in cell regulation motifs close to criticality}

\author{Daniele Proverbio$^{1,2}$} \email{daniele.proverbio@uni.lu}
\author{Arthur N. Montanari$^{1}$} 
\author{Alexander Skupin$^{1,3,4}$} 
\author{Jorge Gonçalves$^{1,5}$}
\affiliation{1- Luxembourg Centre for Systems Biomedicine, University of Luxembourg, 6 Avenue du Swing, 4367, Belvaux, Luxembourg}
\affiliation{2- College of Engineering, Mathematics and Physical Sciences, University of Exeter, EX4 4QL, Exeter, UK}
\affiliation{3- Department of Physics and Material Science, University of Luxembourg, 162a Avenue de la Faiencerie, 1511 Luxembourg, Luxembourg}
\affiliation{4- Department of Neuroscience, University of California San Diego, 9500 Gilman Drive, La Jolla, CA, United States}
\affiliation{5- Department of Plant Sciences, University of Cambridge, CB2 3EA, Cambridge, UK}

\date{\today}
\begin{abstract}

Bistable biological regulatory systems need to cope with stochastic noise to fine-tune their function close to bifurcation points. Here, we study stability properties of this regime in generic systems to demonstrate that cooperative interactions buffer system variability, hampering noise-induced regime shifts. Our analysis also shows that, in the considered cooperativity range, impending regime shifts can be generically detected by statistical early warning signals from distributional data. 
Our generic framework, based on minimal models, can be used to extract robustness and variability properties of more complex models and empirical data close to criticality.

\end{abstract}
%
%
%
%
\maketitle

Many biological systems self-regulate their functions through bistable circuits, which have been associated to genetic \cite{angeli2004detection,kheir2019evolutionary} as well as growth feedbacks \cite{deris2013innate}. In particular, positive feedback loops have long been studied in systems and synthetic biology~\cite{de2016cell,acar2005enhancement,guinn2020observation}; they regulate crucial functions like enzymatic activity or gene transcriptional changes during cell fate decisions~\cite{huang2007bifurcation,fiorentino2020measuring}. Autoactivating positive feedback loops, simple circuit motifs promoting bistability and fine regulation of dynamical states close to self-organised criticality, are of particular importance~\cite{alon2019introduction,Tripathi2020}. Cellular heterogeneity, \textit{i.e.}, random cell-to-cell variations \cite{komin2017address}, can further direct transitions \cite{kaern2005stochasticity,weber2013stochastic} and induce regime shifts between alternative stable states of gene expression or of protein concentrations \cite{thomas2014phenotypic}. Positive feedback loops with stochastic fluctuations have been observed in a variety of system including the transcription network of \textit{E. coli} \cite{milo2002network} or in the regulation of $\beta$\textit{-galactosidase} \cite{ozbudak2004multistability}, which results from a sudden transition from low (``off'') to high (``on'') level states of the \textit{lac operon} at a critical point of an inducer concentration. 

There are mainly two ways in which bistable systems can switch between alternative steady states \cite{ashwin2012tipping}: transitions driven by bifurcations (which, due to loss of system resilience \cite{scheffer2009early}, may be anticipated by small random fluctuations) and transitions driven by large random jumps. Cells and other biological systems are hypothesized to live close to criticality to quickly respond to changing environmental conditions \cite{mora2011biological}, but they should not respond to random environmental changes (noise) in order to maintain their evolutionary fitness.
Close to criticality, the dynamical motifs have reduced resilience and the system can exhibit increasing variability in response to noise \cite{mojtahedi2016cell,sharma2016anticipating}. This is typical of nonlinear systems approaching a critical bifurcation and corresponds to augmented sensitivity to random perturbations and diverging response time, a phenomenon known as critical slowing down (CSD)~\cite{scholz1987nonequilibrium,byrd2019critical}. 

Mechanisms to buffer variability while maintaining the critical state are thus necessary to finely regulate desired transitions \cite{ozbudak2004multistability} or to better cope with undesired shifts \cite{scheffer2009early}. To this end, two main strategies can cooperate: moving the system state away from the bifurcation point, or deepening the basin of attraction to avoid random fluctuations pushing the system state to undesired attractors. These strategies correspond to changes in different environmental or regulatory conditions in cellular systems \cite{dai2015relation}, allowing organisms to exploit different mechanisms to buffer variability close to criticality.
In mono-stable systems, noise can be bound by the action of molecular compounds like microRNAs \cite{siciliano2013mirnas} as well as by temporal relays of signalling molecules \cite{lestas2010fundamental,del2016control}. For critical regimes in bistable processes, a key mechanism to buffer variability is identified here: the cooperative interactions tuning the activation function of positive feedback loops.

\begin{figure}[t]
\centering
\includegraphics[width=.75\linewidth]{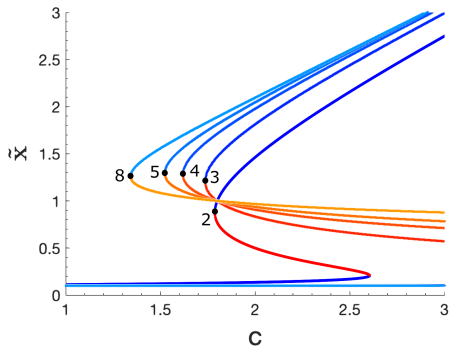}
\caption{\small Bifurcation diagrams $(\tilde{x} ,c)$ of Eq. \eqref{eq:toy_eq} for different Hill coefficients $n$. Stable and unstable branches are represented by blue and red colors, respectively. Black dots identify the bifurcation (saddle-node) points $c_0$. Each value of $n$ is displayed close to its corresponding diagram.}
\label{fig:bif_diag}
\end{figure}

This buffering mechanism can be analysed by considering the simple and well-known adimensional model for stochastic autoactivating positive feedbacks~\cite{santillan2008use,strogatz2018nonlinear}:
\begin{equation}
	\dot x = f(x) + \eta(t) =  K + c \frac{x^n}{1+x^n} - x + \eta(t) \; .
	\label{eq:toy_eq}
\end{equation}
This model describes the Michaelis-Menten kinetics of a transcriptional factor activator (denoted by $x$) \cite{frigola2012asymmetric, weber2013stochastic}, which activates its own transcriptions when bound to a responsive element (Fig.~S1~\cite{supmat}). System \eqref{eq:toy_eq} arises from a model reduction of a two-variables genetic toggle switch, under the assumption of slow-fast timescale separation between the two variables \cite{strogatz2018nonlinear}. Here, $f(x)$ groups the deterministic terms, with steady state $\tilde{x}$ ($\dot{x}|_{\tilde{x}} = 0$), $K$ is the basal expression rate, and $c$ is the maximum production rate with critical value $c_0$ marking a saddle-node bifurcation (Fig. \ref{fig:bif_diag}). The dissociation constant in the denominator of the Hill function was normalised to~1 without loss of generality \cite{smolen1998frequency}. The noise term $\eta(t)$ accounts for intrinsic stochasticity of biological processes \cite{hasty2000noise}. We consider additive Gaussian white noise to approximate the fast degrees of freedom associated to a mean field regime \cite{Berglund2006,sharma2016anticipating}, with the statistical properties $\langle \eta \rangle = 0$, $\langle \eta(t)\eta(t') \rangle = 2 \sigma \delta(t-t')$, where $\sigma$ represents its intensity.  

The Hill coefficient $n$, which describes the nonlinear cooperative binding mechanisms, is usually interpreted as the number of transcription factors that cooperatively promote transcription \cite{santillan2008use}. The smallest value inducing bistability in the circuit is $n = 2$, while $n \to \infty$ yields the logic approximation for the activating Hill function, 
\begin{equation}
   \lim_{n \to \infty}  \frac{x^n}{1+x^n} = \Theta (x-1) \, ,
   \label{eq:logicmodel}
\end{equation}
where $\Theta(\cdot)$ is the Heaviside step function, making system~\eqref{eq:toy_eq} a discrete switch without bistability nor CSD.

This Letter investigates the dependence of resilience properties \cite{dai2015relation} on the cooperativity index $n$. This way, we test how cells can keep production rates \textit{c} close to their critical values and nonetheless increase resilience and buffer variability using other regulation mechanisms. As shown in Fig.~\ref{fig:bif_diag}, increasing $n$ from 2 yields different bifurcation diagrams, where critical points shift to the left and the distance between the upper stable manifold and the unstable manifold decreases when the system gets close to criticality. We focus on systems residing on the upper branch (to be consistent with the mean field assumption) and moving left towards the saddle-node bifurcation point. This way, we investigate how biological circuits can buffer variability close to critical states by exploiting dynamical mechanisms. We also assess the parameter range where CSD-based early warning signals correctly indicate impending regime shifts.

To characterise the system stability properties, we analyse the stationary potentials and probability density functions (PDF) depending on $n$, in analogy to previous works \cite{friedman2006linking,kumar2014exact}. Consider the forward Fokker-Plank equation for the probability density function $P(x,t)$ associated with Eq.~\eqref{eq:toy_eq}:

\begin{equation}
    \frac{\partial P(x,t)}{\partial t} = -\frac{\partial}{\partial x} \left[f(x)P(x,t) \right] + \frac{\partial^2}{\partial x^2} \left[\sigma P(x,t) \right] \, ,
\end{equation}
where $f(x)$ lumps the deterministic terms of Eq.~\eqref{eq:toy_eq}. The stationary solution $P_s(x)$ takes the form \cite{Gardiner1985}
\begin{align}
    P_s(x) & = N_c e^{-\phi(x)} \, ,\\
    \phi(x)  & = \frac{1}{2}\ln \sigma - \frac{1}{\sigma}\int^x f(x')dx' \, ,
\end{align}

\noindent
where $\phi(x)$ describes the adjoint stochastic potential whose depth is related to system resilience, \textit{i.e.}, its ability to recover after a perturbation.  $N_c$ is a normalization constant such that $\int_\Omega P_s(x)=1$ ($\Omega$ is the domain).

\begin{figure}[t]
\centering
\includegraphics[width=\linewidth]{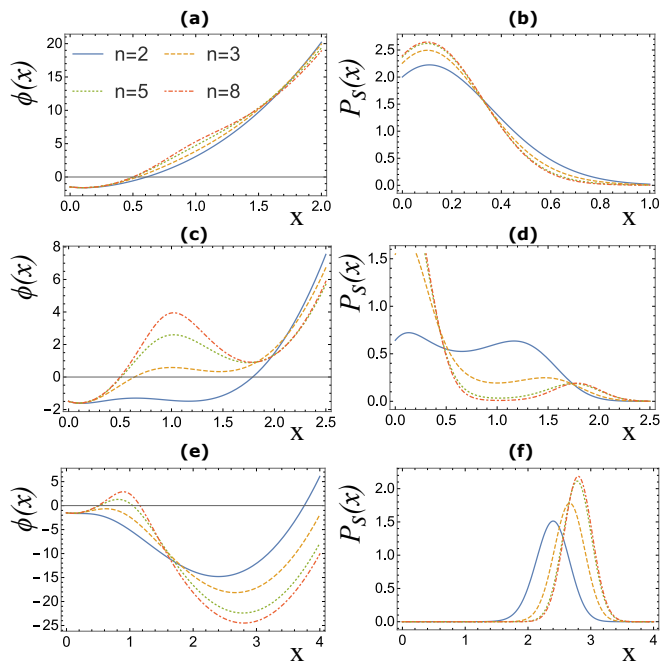}
\caption{\small Effect of the Hill coefficient $n$ on the stochastic potential $\phi(x)$ (left column) and on the stationary probability density function $P_s(x)$ (right column), when only additive white noise is present. (a, b) $c=0.8$ (``off'' state); (c, d) $c = c_0(n) + 0.05$ (multistable region); (e, f) $c = 2.7$ (``on'' state favoured). In all cases, $K=0.1$ and $\sigma = 0.05$. }
\label{fig:FP}
\end{figure}

Fig. \ref{fig:FP} shows the dependency of $\phi(x)$ and $P_s(x)$ on $n$ when the system is either in an ``off'' state far away from criticality (Fig.~\ref{fig:FP}a,b), close to the criticality (Fig.~\ref{fig:FP}c,d) or beyond it, where the ``on'' state is favoured (Fig.~\ref{fig:FP}e,f). Increasing the cooperativity index $n$ does not alter the underlying bistability, but modifies the depth of the potential and increases the separation of alternative states (Fig.~S2 \cite{supmat}). 
For the ``off'' and ``on'' states, the corresponding equilibria exhibit  significantly deep attractor basins with only minor dependence on $n$, as also indicated in the bifurcation diagrams (Fig. \ref{fig:bif_diag}) and by $P_s(x)$. Close to critical points, the picture changes. The potential $\phi(x)$ displays two commensurable wells, which are more evident and symmetric for larger $n$, suggesting that both states become equally occupied in noisy environments. For increasing $n$, $P_s(x)$ displays sharper peak separation between the bistable states: the system diffuses less to intermediate states and is more constrained around single equilibrium values, as anticipated due the steeper potential barriers in $\phi(x)$. Random deviations are thus suppressed faster and transitions from one state to another are sharper and thus more robust against noise.

We now focus on how $n$ influences variability measures, like variance and autocorrelation, close to criticality. Obtaining globally analytic expressions is challenging, in particular for high values of $n$. Hence, we focus on a local analysis close to the bifurcation points and employ a geometrical methodology. To derive generic results for critical manifolds, we use their local topological equivalence to bifurcation normal forms \cite{kuehn2021universal}. The normal forms associated to dynamical systems are simplified minimal-order forms to which all systems exhibiting a certain type of bifurcation are, around the equilibrium, topologically equivalent \cite{kuznetsov2013elements}. Supplementary Material \cite{supmat} provides a brief background to normal forms and terminology used in this work.
For saddle-node bifurcations like in Fig.~\ref{fig:bif_diag}, the associated normal form is \cite{kuznetsov2013elements}
\begin{equation}
    \dot{x} = p - x^2 \, ,
    \label{eq:normalform}
\end{equation}
with two equilibrium manifolds $\tilde{x}_{1,2} = \pm \sqrt{p}$, one stable ($+$) and the other unstable ($-$). Note that the normal form corresponds to a parabola. To study the behaviour of stochastic solutions near the stable manifold, consider the evolution of its first-order perturbation, $y = \delta x|_{\tilde{x}_1}$ exposed to the same additive white noise $\eta(t)$ as in Eq.~\eqref{eq:toy_eq} \cite{kuehn2011mathematical}. Since $k = 2\sqrt{p}$ is the distance of the control parameter value from its critical value $p_0 = 0$, note that $k$ is proportional to $c-c_0$ from the original system, following normal form properties. The corresponding Langevin equation accounting for mean field fluctuations around the stable equilibrium is then given by
\begin{equation}
    \dot{y} = -ky + \eta(t) \, .
    \label{eq:OU}
\end{equation}
Eq.~\eqref{eq:OU} is a typical Ornstein–Uhlenbeck (OU) process with exact solutions for statistical moments~\cite{Gardiner1985}.

To connect the quantitative effects of $n$ with the more qualitative topological form Eq.~\eqref{eq:normalform}, recall that $n$ widens or narrows the local parabolic shape of the original bifurcation diagram for Eq.~\eqref{eq:toy_eq} (Fig.~\ref{fig:bif_diag}). Eq.~\eqref{eq:normalform} thus needs to be augmented with a term $\rho$ to modify the focal width of its parabolic stable manifold, which corresponds to the width of the parabola at the focal point. This leads to
\begin{equation}
    \dot{x} = p - \rho x^2 \, .
    \label{eq:new_normal_form}
\end{equation}
In this formulation, $\rho$ corresponds to the focal width of the normal form. Supplementary Material \cite{supmat} contains analytical derivations for the approximation of system \eqref{eq:toy_eq} to the normal form \eqref{eq:new_normal_form}, and its relationships with the geometrical results. Propagating $\rho$ into Eq.~\eqref{eq:OU} adds a tuning term to the bifurcation parameter, $k \to \sqrt{\rho} k$. Hence, the corresponding OU process for a semi-quantitative saddle-node normal form is
\begin{equation}
    \dot{y} = -\sqrt{\rho}ky + \eta(t) \, .
    \label{eq:OUsemiquant}
\end{equation}

Among its statistical moments and power spectral properties, we are primarily interested in quasi-steady-state variance ($\text{Var}$) and lag-1 autocorrelation ($\text{AC1}$), measures of system variability close to criticality. They have been proposed as proxies for system resilience and early warning signals (EWS) of impending bifurcation points \cite{scheffer2009early,Trefois2015a}. 
Based on our mapping to the OU process (Eq.~\eqref{eq:OUsemiquant}), the analytical solutions for $\text{Var}$ and $\text{AC1}$ take the form \cite{Gardiner1985}:
\begin{equation}
    \text{Var}  = \frac{\sigma}{\sqrt{\rho}k}, \quad \text{AC1}   = e^{-\sqrt{\rho} k} \, . \label{eq:var}
\end{equation}

\begin{figure}[t]
\centering
\includegraphics[width=\linewidth]{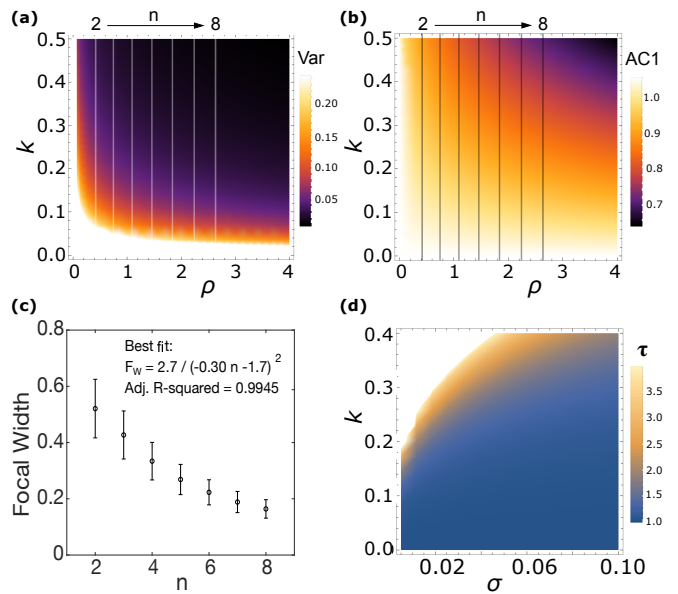}
\caption{\small Theoretical dependence of EWS measures (a) $\text{Var}$ and (b) $\text{AC1}$ on $\rho$ (related to focal width) and $k$ (distance measure from critical parameter values). Vertical lines represent slices for fixed values of $\hat{\rho}$ (Eq.~\eqref{eq:rho_emp}) corresponding to the mean $F_W$ shown in panel~c. The associated $n$ values increase from left to right. (c) Relationship between $F_W$ (Eq. \eqref{eq:FW}) and corresponding values of $n$, with best fit. Error bars correspond to one standard deviation (Eq.~\eqref{eq:err_prop}). (d) Generic escape rate $\tau$ as a function of noise level $\sigma$ and $k$ (Eq.~\eqref{eq:kramers}).}
\label{fig:var_ac3}
\end{figure}

Eqs.~\eqref{eq:var} are generic for noisy saddle-node bifurcations. Fig.~\ref{fig:var_ac3}a,b shows them as functions of $\rho$ and $k$. To connect with the original autoactivating feedback system, we estimate the focal width of the bifurcation diagrams for each $n$ by fitting a parabolic form $c = \alpha \tilde{x}^2 + \beta \tilde{x} + \gamma$ to the data points of each bifurcation diagram in the vicinity of the saddle point. Using \textsc{Matlab} Curve Fitting toolbox also provides uncertainties over $\vec{\theta} = [\alpha, \beta, \gamma ]$, resulting from small deviations from a perfect parabolic shape. By definition, the fitted focal width is
\begin{equation}
    F_W = 2|\tilde{x}(c_F) - \tilde{x}_F| \, ,
    \label{eq:FW}
\end{equation}
where ($\tilde{x}_F$, $c_F$) are the coordinates of the parabolic focus. To get a reasonable estimate of the corresponding uncertainties, the associated standard deviation is derived from the fitted parameter uncertainties $\text{std}(\bar{\theta}_i)$ using a first-order approximated propagation method \cite{taylor1997introduction}:
\begin{equation}
    \text{std}(F_W) = \left[ \sum_i \left(\frac{\partial F_W}{\partial \theta_i} \text{std}(\theta_i)  \right)^2 \right]^{\frac{1}{2}} \, .
    \label{eq:err_prop}
\end{equation}

The relationship between $F_W$ and $n$ is plotted in Fig.~\ref{fig:var_ac3}c, with the corresponding $\text{std}(F_W)$. The pattern decreases quadratically, thereby marking a rapid decrease followed by almost plateauing. Hence, a bounded and relatively small cooperativity index is in principle sufficient to effectively buffer variability close to criticality.

The estimated $\hat{\rho}$ values from fitted focal widths, for $n=2$ to $n=8$, are obtained as 
\begin{equation}
    \hat{\rho} = \xi (F_W)^{-1} \, ,
    \label{eq:rho_emp}
\end{equation}
where $\xi$ is a tuning parameter proportional to the Hill function (Supplementary Material \cite{supmat}).
Mean $\hat{\rho}$ values are marked in Fig. \ref{fig:var_ac3}a,b with solid vertical lines. Consistently with the trend observed in Fig. \ref{fig:var_ac3}c, the mean values spread as $n$ increases (from left to right). Low $n$ values yield higher sensitivity to noise, as both Var and AC1 show substantially higher values for small cooperativity indices $n$, even when $k$ is large (\textit{i.e.}, further away from the critical point, but still within the bistable region, \textit{cf.} Fig. \ref{fig:bif_diag}). Thus, values of $\rho$ can belong to two regions: one, where the values for both metrics are high for all $k$ (left side of Fig. \ref{fig:var_ac3}a,b), or another one where both metrics maintain low values for most $k$ and increase rapidly close to criticality (right side of Fig. \ref{fig:var_ac3}a,b). The region $\hat\rho \to \infty$ corresponds to the logic approximation \eqref{eq:logicmodel} with $n \to \infty$,  where $\text{Var}$ and AC1 also change abruptly in a step-wise manner. The ultra-sensitive region $\hat\rho \to 0$ is spanned by increasing dissociation constants (Fig.~S3 \cite{supmat}), and potentially by changing other parameters, here not explicitly considered, or by different activation functions describing, for example, wild-type vs mutant organisms~\cite{hasty2000noise}. Other pathways like growth feedbacks \cite{deris2013innate} will likely correspond to additional regions in the parameter space. These investigations are left to future studies.

\begin{figure}[t]
\centering
\includegraphics[width=\linewidth]{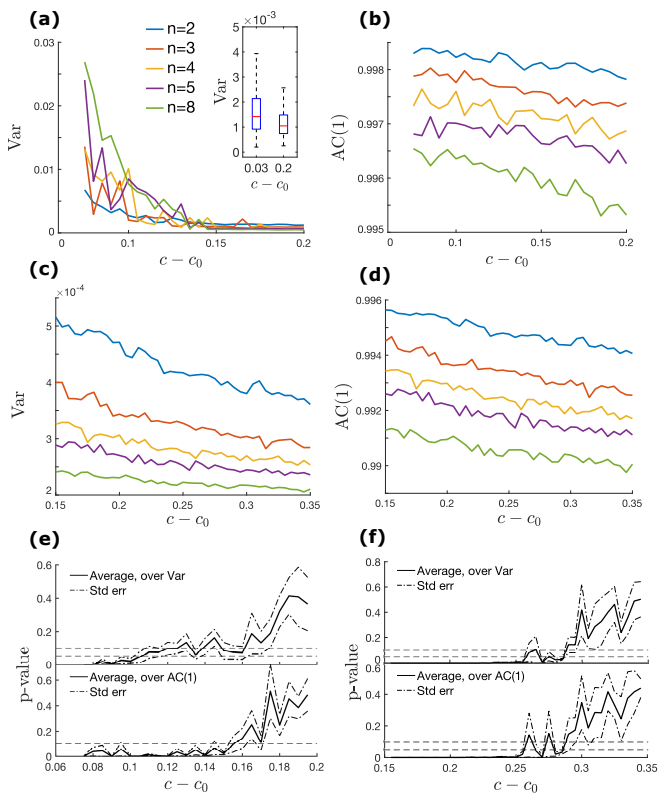}
\caption{\small Average trends of (a) $\text{Var}$ and (b) $\text{AC1}$ as a function of $c$ close to bifurcation $c_0$, in the regime where noise-induced transitions might occur (blue area of Fig.~\ref{fig:var_ac3}d). Simulations are presented over 200 realizations for different $n$. For each $c$, the indicators spread into distributions, as exemplified for two $c$ values in inset (a). (c,d) Average trends of $\text{Var}$ and $\text{AC1}$ farther from bifurcation $c_0$. (e,f) Evolution of p-values between $\text{Var}$ and $\text{AC1}$ distributions at each $c-c_0$ and the ``reference'' distribution. The ``reference'' distribution corresponds to (e) $c-c_0 = 0.2$ (starting ``closer to bifurcation'') and (f) $c-c_0 = 0.35$ (``farther from bifurcation''). Dashed lines represent typically used p-values in biological experiments.}
\label{fig:sims}
\end{figure}

We finally investigate the performance of EWS against impending bifurcation points. The motivation is the following: consider complex systems lacking validated mechanistic models; in our case, this would translate to a scenario where  $n$\textemdash or even the precise activation function\textemdash of an eukaryotic cell is poorly identifiable~\cite{hasty2000noise}. This consideration leads to questioning if we can identify statistical signals, computed on empirical data, that provide reliable information about the system's loss of resilience. Increasing trends of $\text{Var}$ and $\text{AC1}$ have been widely suggested to work as EWS \cite{scheffer2009early,Trefois2015a} but their robustness remains elusive. To study how generic they are in the identified parameter range and to account for mean trends and uncertainties, we numerically integrate the original stochastic system \eqref{eq:toy_eq} using the Euler-Maruyama method. To mimic cell populations slowly evolving close to equilibrium, we sample $10^4$ time points over 200 repeated experiments in dependence of $c$. This leads to a distribution of statistical indicators (\textit{e.g.}, see Fig.~\ref{fig:sims}a, inset).

To distinguish between bifurcation-induced transitions, anticipated by loss of resilience, and noise-induced transitions, we measure the scale between the distance to the bifurcation point and the noise level by the Kramers escape rate $\tau = 2\pi(\sqrt{|U''(\tilde{x}_1)U''(\tilde{x}_2)|})^{-1}\exp[(U(\tilde{x}_2)-U(\tilde{x}_1))/\sigma]$ \cite{Gardiner1985}. For any saddle-node bifurcation manifold \eqref{eq:new_normal_form} equipped with additive noise, $U(\tilde x_2) - U(\tilde x_1) = (4/3)(k^3 /\sqrt{\rho})$ and $|U''(\tilde x_{1,2})| = 2\sqrt{p\rho}$. Hence,
\begin{equation}
    \tau = \frac{\pi}{k\sqrt{\rho}}\exp\left[\frac{4}{3}\frac{k^3}{\sqrt{\rho}\sigma} \right] \, .
\end{equation}
Values lying at the exponential boundary of
\begin{equation}
    \tau \simeq \mathcal{O}(\exp[k^3/\sigma]) \,
    \label{eq:kramers}
\end{equation}
(Fig.~\ref{fig:var_ac3}d) provide comparable ranges of control parameters and noise levels for all simulations with different $n$. They distinguish two regimes, one where few noise-induced transitions might occur ($\tau \lesssim 2$, Fig.~\ref{fig:sims}a,b) and another regime primarily determined by bifurcation-driven resilience loss ($\tau \gtrsim 2$, Fig.~\ref{fig:sims}c,d). For the considered $\sigma = 0.02$, the system is very close to critical points.

Fig.~\ref{fig:sims}a--d display average values for $\text{Var}$ and $\text{AC1}$ from numerical simulations. When the dynamics is mostly characterised by the bifurcation (Fig.~\ref{fig:sims}c,d), both measures display patterns consistent with those predicted in Fig.~\ref{fig:var_ac3}a,b and increasing $n$ better buffers variability. When the noise level becomes comparable to the potential depths, $\text{Var}$s for different $n$ become very close to each other due to the more prominent role of noise-induced uncertainties (Fig. \ref{fig:sims}a). By contrast, $\text{AC1}$s (Fig.~\ref{fig:sims}b) remain separated due to their lower sensitivity to noise (\textit{cf.} Eq. \eqref{eq:var}), but with less marked\textemdash and, therefore, harder-to-detect\textemdash trends, similarly to those observed in real-world data \cite{proverbio2022performance}.

For online applications (\textit{i.e.,} as new data come in and without future knowledge of the system evolution), it is necessary to quantify whether an observed increasing trend is statistically significant, assessing whether it corresponds to a EWS or some spurious fluctuation \cite{boettiger2012early}. To do so, we look for significant p-values between the computed distributions close to criticality and those far from the bifurcation (``reference'') (inset in Fig.~\ref{fig:sims}a). Fig.~\ref{fig:sims}e,f show patterns of p-values from $\text{Var}$ and $\text{AC1}$, averaged over all $n$. The p-values are computed closer to the bifurcation in Fig.~\ref{fig:sims}e ($c-c_0$ corresponding to the parameters in Fig.~\ref{fig:sims}a,b) and farther from the bifurcation in Fig.~\ref{fig:sims}f ($c-c_0$ corresponding to the parameters in Fig.~\ref{fig:sims}c,d). The p-values cross their significance levels (either 0.1 or 0.05 \cite{andrade2019p}, dashed lines in Fig.~\ref{fig:sims}e,f) before the bifurcation point. This assesses that significant increasing trends of statistical indicators can be detected prior to the transitions, thus constituting reliable early warning signals. This analysis thus certifies the potential use of proposed EWS to detect approaching bifurcation points in biological motifs, providing a quantification of how much in advance the EWS become significant depending on the reference and on the p-values threshold. \\


Overall, our study characterised fundamental dynamical mechanisms to buffer systems' variability in critical regimes. We determined parameter ranges, corresponding to plausible cooperativity values for the positive feedback loop motif, where both variance and autocorrelation display low relative sensitivity to additive noise. In other ranges, however, the system poorly buffers its variability. Investigating whether these ranges could correspond to other dynamical mechanisms is demanded for future studies. Moreover, state-dependent noise can be further incorporated in model \eqref{eq:toy_eq} to make it closer to biological reality \cite{hasty2000noise}. Although it was not explicitly considered in this paper, primarily to focus on the effects of a single parameter on the system's stability properties, in Supplementary Material \cite{supmat} (also including Refs \cite{bruggeman2018living,Das2010,ZAHRI2014186}) we investigate its influence on the variability metrics considered above. Notably, it does not alter trends of AC(1) but affect those of Var (Fig. S4 \cite{supmat}), as expected from dependencies on $\sigma$ in Eq. \eqref{eq:var}. This is relevant in further buffering fluctuation amplitudes near criticality and calls for caution when using Var as an EWS indicator in systems strongly characterised by such type of noise. A deep investigation of interplays between noise types and resilience properties is left for future works. These may unravel alternative ways by which cells regulate their states or support the hypothesis of a self-organised fine-tuning in ``safe'' parameter spaces. Overall, our analysis contributes with quantitative insights to analytical and experimental studies of bistable systems' resilience and connects general and system-specific predictions \cite{dai2015relation}.

We also assessed the sensitivity of proposed EWS to an additional regulation mechanism and suggested how to on-line quantify significant increasing trends from distributional data. We showed that extra parameters in the dynamical system do not alter the warning capabilities of indicators associated with CSD. In the considered parameters' range, they are sufficiently generic to detect resilience loss.  As several indicators have been developed upon, to detect cell-fate decisions \cite{mojtahedi2016cell} and to possibly anticipate undesired shifts to, \textit{e.g.}, cancerous states \cite{yang2018dynamic,aihara2022dynamical}, our results constitute an important step to interpret and apply them correctly. However, their use should be treated carefully if other quantitative mechanisms between noise and bifurcations might be at play, like certain types of state-dependent noise, possibly shadowing theoretical trends. Following our methodology, future studies might inquire other indicators and their behaviour under changing $n$ and additional conditions. Our framework can also be easily extended to inquire the performance of buffers and EWS in different dynamical models and experimental setups. \\



%
%
%
%
\textit{Acknowledgements} The authors thank J. Fuentes for valuable discussions. D.P. is supported by the Luxembourg National Research Fund (FNR) PRIDE DTU CriTiCS (10907093) and A.S. by the FNR (C14/BM/7975668/CaSCAD) and by the NIH NBCR (NIH P41 GM103426).

%
%
%
%


%

\end{document}